\documentclass[aps,twocolumn,showpacs,prl,10pt,superscriptaddress,preprintnumbers,nofootinbib]{revtex4-2}
\usepackage{amsmath, amssymb}
\usepackage{physics, bm}
\usepackage{graphicx}
\usepackage{xcolor}
\usepackage[colorlinks, linkcolor=red, citecolor=red, urlcolor=magenta]{hyperref}

\usepackage[normalem]{ulem}
\usepackage{soul}
\usepackage{multirow}
\usepackage{orcidlink}

\newcommand{\beq}[1][]{\begin{equation}\label{#1}}
\newcommand{\eeq}{\end{equation}}
\newcommand{\nn}{\nonumber}

\begin{document}

\preprint{
	{\vbox {
		\hbox{\bf CPTNP-2026-010}
}}}
\vspace*{0.2cm}

\title{Probing the Color-Octet Mechanism via Dihadron Fragmentation in $\chi_b$ Decays}

\author{Zhi-Guo He\, \orcidlink{0000-0001-9887-2058}}
\email{zhiguo.he@buct.edu.cn}
\affiliation{Department of Physics and Electronics, School of Mathematics and Physics, Beijing University of Chemical Technology, Beijing 100029, China}

\author{Guanghui Li\,\orcidlink{0009-0001-4822-3321}}
\email{ghli@ihep.ac.cn}
\affiliation{Institute of High Energy Physics, Chinese Academy of Sciences, Beijing 100049, China}
\affiliation{School of Physical Sciences, University of Chinese Academy of Sciences, Beijing 100049, China}

\author{Yu-Jie Tian\,\orcidlink{0009-0002-7071-0916}}
\email{yjtian@ihep.ac.cn}
\affiliation{Institute of High Energy Physics, Chinese Academy of Sciences, Beijing 100049, China}
\affiliation{School of Physical Sciences, University of Chinese Academy of Sciences, Beijing 100049, China}

\author{Xin-Kai Wen\,\orcidlink{0009-0008-2443-5320}}
\email{xinkaiwen@ihep.ac.cn (corresponding author)}
\affiliation{Institute of High Energy Physics, Chinese Academy of Sciences, Beijing 100049, China}
\affiliation{China Center of Advanced Science and Technology, Beijing 100190, China}

\author{Bin Yan\,\orcidlink{0000-0001-7515-6649}}
\email{yanbin@ihep.ac.cn (corresponding author)}
\affiliation{Institute of High Energy Physics, Chinese Academy of Sciences, Beijing 100049, China}
\affiliation{Center for High Energy Physics, Peking University, Beijing 100871, China}

\date{\today}
\begin{abstract}
The color-octet (CO) mechanism is a cornerstone of nonrelativistic QCD, yet knowledge of its long-distance matrix elements remains limited, preventing stringent tests of the theory. We demonstrate that the Artru-Collins asymmetry in hadronic decays of the $P$-wave bottomonium state $\chi_{b2}$ provides a direct probe of CO dynamics. At leading order, the asymmetry arises exclusively from the CO decay channel, whereas the color-singlet (CS) contribution affects only the unpolarized rate, so that a nonzero signal constitutes unambiguous evidence of the CO mechanism. This observable provides a novel way to extract the ratio $\rho_8$ between CO and CS matrix elements. Focusing on $e^+e^-\to\Upsilon(2S)\to\gamma\,\chi_{b2}$ at Belle, we show that the asymmetric beam configuration preserves the asymmetry in the laboratory frame and avoids the strong suppression present in the center-of-mass frame. With the Belle II dataset, $\rho_8$ could be determined with sufficient precision to address the long-standing discrepancy between the lattice calculations and phenomenological determinations.
\end{abstract}

\maketitle

\emph{Introduction---}
Five decades after the discovery of the $J/\psi$ meson, heavy quarkonium 
production and decay processes remain a preeminent laboratory for probing the 
interplay between perturbative and nonperturbative regimes of quantum chromodynamics (QCD). The phenomenology of their spectrum, decay, and 
production facilitates the extraction of fundamental standard model parameters, 
such as the heavy quark mass $m_{Q}$ and the strong coupling constant 
$\alpha_s(m_Q)$. Grounded in rigorous effective field 
theory~\cite{Lepage:1992tx}, the nonrelativistic QCD (NRQCD) 
factorization~\cite{Bodwin:1994jh} is now the most robust framework for describing 
heavy quarkonium inclusive decay, in which the decay width is factorized into 
short-distance coefficients (SDCs) to describe the annihilation of the heavy-quark pair ($Q\bar{Q}$) into light partons
and nonperturbative long-distance matrix elements (LDMEs) to characterize 
the probability of that $Q\bar{Q}$ pair inside the meson. In contrast to the 
traditional color-singlet model, NRQCD factorization allows the 
intermediate $Q\bar{Q}$ pair to be in all possible spin and color 
configurations, particularly in the color-octet (CO) state known as the CO 
mechanism (COM). The introduction of COM is not an ad hoc supplement, but rather 
an inevitable consequence of QCD factorization and NRQCD velocity power 
counting~\cite{Lepage:1992tx}. It is essential for ensuring infrared safety, 
restoring factorization to all orders of $\alpha_s$ in perturbative calculation. 
Consequently, a precise determination of the CO LDMEs provides definitive 
evidence of NRQCD factorization in the decay sector. 

The $P$-wave quarkonium $\chi_{QJ}$ decay is viewed as the golden channel to 
probe COM because both color singlet (CS) ${}^3P_J^{[1]}$ and CO $^3S_1^{[8]}$ 
states contribute at $v^2$ leading order (LO) in NRQCD 
factorization~\cite{Bodwin:1994jh}. 
While the CS LDMEs 
$H^{Q}_1=\langle \chi_{QJ}|\mathcal{O}(^3P_J^{[1]})|\chi_{QJ} \rangle$ can be 
related to the first derivative of the wave function at the origin within 
potential model calculations, the CO ones 
$H_8^{Q}(\mu_\Lambda)=\langle \chi_{QJ}|\mathcal{O}(^3S_1^{[8]},\mu_\Lambda)|\chi_{QJ} \rangle$, 
where $\mu_\Lambda$ is the NRQCD factorization scale, have to be determined via 
simulations of lattice NRQCD or fits to experimental data~\cite{Bodwin:1992ye,Bodwin:2001mk}. In 
the charmonium case, the LDMEs predicted by lattice NRQCD~\cite{Bodwin:1996tg} are 
in good agreement with the fits to $\chi_{cJ}$ 
decay~\cite{Maltoni:2000km,Bodwin:2005ec}. However, for the bottomonium states 
$\chi_{bJ}(1P)$, the lattice prediction for the ratio 
$\rho_8(m_b)=H_8^{b}(m_b)m_b^2/H_1^{b}=0.044\pm 0.015$~\cite{Bodwin:2001mk,Bodwin:2007zf} is 
significantly smaller than the value extracted from experimental fits to $D^{0}$ production in $\chi_{bJ}(1P)$ 
decays, $\rho_8(m_b)=0.16^{+0.071}_{-0.047}$~\cite{CLEO:2008bsq}. 
Furthermore, $\rho_8(m_b)$ can also be related to the gluonic correlator $\mathcal{E}_3$ 
in potential-NRQCD (pNRQCD), which was fitted to $\chi_{cJ}$ decay 
data~\cite{Brambilla:2001xy,Brambilla:2020xod} or computed in the refined Gribov-Zwanziger theory~\cite{Chung:2023bjr}. 
An independent determination of $\rho_8(m_b)$ from $\chi_{bJ}$ decay is essential to examining the universal nature 
of $\mathcal{E}_3$ in strongly coupled pNRQCD. 

In fits to decay rates, the CS and CO contributions are experimentally 
indistinguishable, precluding a precise determination of $\rho_8(m_b)$. 
In this Letter, we propose a novel approach utilizing the transverse spin correlations 
of quarks, described by the Artru-Collins asymmetry~\cite{Artru:1995zu}, as a 
direct probe of the COM in $\chi_{bJ}$ decays at lepton colliders. Because of 
$J^{PC}$ conservation, at QCD LO, the CO $b\bar{b}(^3S_1^{[8]})$ state decays 
into $q\bar{q}$ whereas CS $b\bar{b}(^3P_{J}^{[1]})$ decays into gluons. Both 
the light quarks and gluons will subsequently fragment into dihadron pairs. 
Different from inclusive decay rates, the Artru-Collins asymmetry is sensitive 
to the spin structures of the fragmenting partons~\cite{Artru:1995zu}, thereby 
directly encoding the quantum numbers of the intermediate $q\bar{q}$ or 
gluons. We will demonstrate for the first time that CO and CS channels produce 
distinct dihadron azimuthal asymmetry patterns, enabling the isolation of the CO 
contribution through parton spin correlations in $\chi_{bJ}$ decays.

\vspace{3mm}
\emph{Kinematics and observable---}
At KEKB, $\chi_{bJ}$ states are produced predominantly via $E1$ transitions 
from the transversely polarized $\Upsilon(2S)$ in $e^{+}e^{-}$ annihilation. 
In the heavy-quark limit, the $\chi_{bJ}$ polarization is completely determined by the $E1$ transition amplitude
and directly reflects the parent $\Upsilon(2S)$ polarization~\cite{Eichten:2007qx}.
For $\chi_{b0}$, only the $b\bar b(^3P_0^{[1]})\to gg$ generates an Artru-Collins type asymmetry,
while the CO $b\bar b(^3S_1^{[8]})\to q\bar q$ channel yields none due to the scalar nature of the $\chi_{b0}$. 
For $\chi_{b1}$, the two-gluon channel is forbidden by the Landau-Yang theorem, so the decay is CO dominated, consistent with the observed branching fraction of $\chi_{b1}\to D^0X$~\cite{ParticleDataGroup:2024cfk}. 
The resulting asymmetry is nevertheless insensitive to the LDMEs since they 
enter the polarized and unpolarized cross sections identically and cancel in the 
asymmetry ratio. By contrast, for $\chi_{b2}$ decay the CO channel generates nontrivial transverse  $q\bar q$
spin correlations. Although individual $\chi_{b2}$ helicity states can generate a nonzero linear-polarization correlation
between the two gluons in the CS $gg$ channel, these contributions cancel exactly after summing over the helicity weighted by the $E1$ transition, which is a consequence of the spin-2 angular-momentum structure. Consequently, at LO, the azimuthal asymmetry
is generated by the CO $q\bar q$ channel, while the CS channel contributes only to the unpolarized cross section and dilutes the signal.
We thus focus on $\chi_{b2}\to q\bar q,gg\to (\pi^+\pi^-)+(\pi^+\pi^-)+X$, where the transverse $q\bar q$ spin correlations generate an
asymmetry directly sensitive to the underlying CO dynamics.

\begin{figure}
    \centering
    \hspace*{-0.7cm}
    \includegraphics[width=1.1\linewidth]{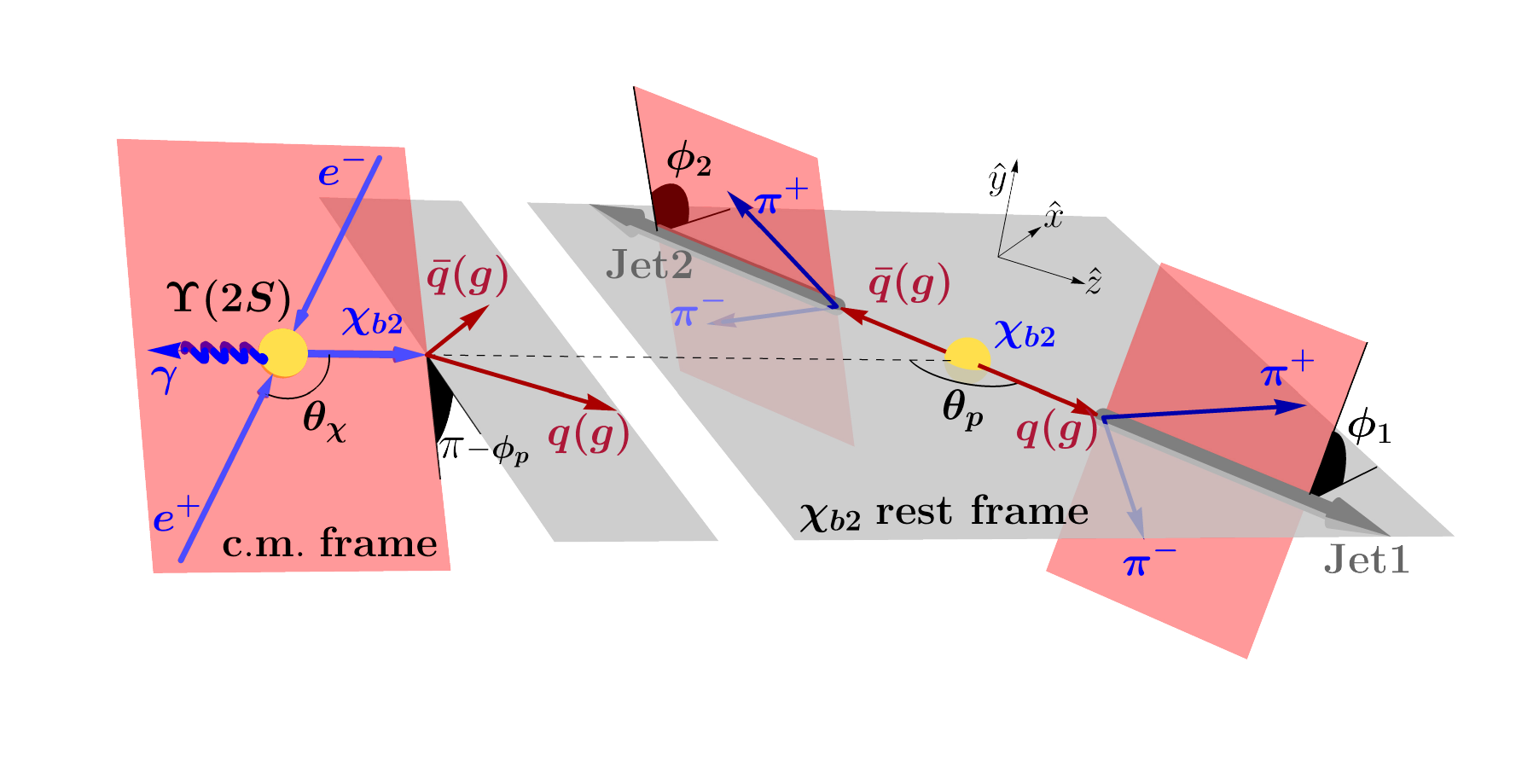}
    \caption{Leading order kinematic configuration for the production of $\pi^+\pi^-$-dihadron pairs in $\chi_{b2}$ decay at lepton colliders.  The $\chi_{b2}$ production is defined in the c.m.~frame, while the partonic subprocess is described in the $\chi_{b2}$ rest frame.}
    \label{fig:geometry}
\end{figure}

Figure~\ref{fig:geometry} illustrates the LO kinematics of this process. The $\chi_{b2}$ production is defined in the center-of-mass (c.m.)~frame, while the subsequent partonic decay and its kinematics are described in the $\chi_{b2}$ rest frame. To characterize the dihadron kinematics, we construct a local coordinate system where the $\hat{z}$ axis is chosen along the thrust direction. The thrust axis, reconstructed from the final-state hadrons, approximates the parton direction. Each $\pi^+\pi^-$ pair is characterized by the total momentum $P_i = p_i^{\pi^+} + p_i^{\pi^-}$ and the relative momentum $R_i = (p_i^{\pi^+} - p_i^{\pi^-})/2$, with $i = 1,2$. Experimentally, the two $\pi^+\pi^-$ pairs can be selected from opposite hemispheres in large-thrust events, thereby enhancing the back-to-back two-jet topology in the $\chi_{b2}$ rest frame and suppressing multijet contamination~\cite{Belle:2011cur}. Within the collinear factorization framework~\cite{Collins:1993kq,Collins:2011zzd} and in the limit $|\bm{P}_i| \gg M_i$, the differential cross section for the $\pi^+\pi^-$-dihadron pair final states can be expressed in terms of the invariant masses $M_i^2 \equiv P_i^2$, the azimuthal angles $\phi_i$ of the transverse component $\bm{R}_{i,T}$, and the kinematic variables $z_i=2P_i\cdot p_\chi/m_\chi^2$, where $p_\chi$ and $m_\chi$ denote the momentum and mass of $\chi_{b2}$, respectively.

Within this framework, the differential distribution of $\pi^+\pi^-$ pairs from $\chi_{b2}$ decay is obtained by combining the radiative production $\Upsilon(2S)\to\gamma \, \chi_{b2}$ in the c.m.~frame with the partonic decay $\chi_{b2}\to q\bar q$ or $gg$ in the $\chi_{b2}$ rest frame,
\begin{align}
&\frac{d\sigma}{\sigma_0dz_1dz_2dM_1dM_2d\phi_1d\phi_2 d\cos\theta_\chi d\cos\theta_p d\phi_p}\nn\\
&=H_8^{b} \sum_q \mathcal{C}_q \, \big[ \, D_1^q(z_1,M_1)D_1^{\bar q}(z_2,M_2) \nn\\
&+\frac{1}{2}\mathcal{B} \, H_1^{\sphericalangle, q}(z_1,M_1)H_1^{\sphericalangle,\bar q}(z_2,M_2)\cos(\phi_1+\phi_2) \,\big]\nn\\
&+ H_1^{b} \mathcal{C}_g \, D_1^g(z_1,M_1)D_1^g(z_2,M_2),
\label{eq:fac}
\end{align}
where $\sigma_0$ denotes the production cross section of $\chi_{b2}$. The angle 
$\theta_\chi$ is the polar angle between the incoming electron and the produced 
$\chi_{b2}$ in the  c.m.~frame. The angles $\theta_p$ and $\phi_p$ are the polar 
and azimuthal angles of the partons in the $\chi_{b2}$ rest frame, defined 
relative to the $\chi_{b2}$ momentum direction in the c.m.~frame (see 
Fig.~\ref{fig:geometry}). The hard coefficients $\mathcal{C}_{q/g}$ are 
determined by the unpolarized decay rates for $\chi_{b2}\to q\bar {q}/ g g$, 
respectively, in which the LDMEs $H_{1,8}^b$ are factorized out and the NRQCD factorization scale is chosen to 
be $m_b$. 

The hadronization process is described by the dihadron fragmentation functions (DiFFs), which encode the probability for a parton to fragment into a specific dihadron final state~\cite{Artru:1995zu,Jaffe:1997hf,Jaffe:1997pv,Bianconi:1999cd,Bianconi:1999uc,Bacchetta:2003vn,Bacchetta:2008wb,Zhou:2011ba,Courtoy:2012ry,Radici:2015mwa,Pitonyak:2023gjx,Cocuzza:2023oam,Cocuzza:2023vqs,Rogers:2024nhb,Pitonyak:2025lin,Mahaut:2025hie,Rogers:2026jca}.
In Eq.~\eqref{eq:fac}, $D_1^q$ and $H_1^{\sphericalangle,q }$  denote the unpolarized and interference DiFFs for quarks, respectively, while $D_1^g$ is the unpolarized gluon DiFF.
We adopt the definitions and normalization convention used in the JAM analyses~\cite{Pitonyak:2023gjx,Cocuzza:2023oam,Cocuzza:2023vqs}.
The functions $D_1^{q,g}$ determine the overall production rates, whereas the chiral-odd $H_1^{\sphericalangle,q }$ probes transverse quark polarization.  All DiFFs depend on the factorization scale $\mu$, which we take as $\mu=m_\chi$, and the scale dependence is suppressed here for simplicity.  Owing to this chiral-odd structure, DiFFs provide a clean probe of transverse spin dynamics of quarks and offer a sensitive avenue to search for potential new physics~\cite{Zhou:2011ba,Yang:2024kjn,Wen:2024cfu,Wen:2024nff,Cheng:2025cuv,Cao:2025qua,Cao:2025wfg}. 
The coefficient $\mathcal{B}=C_{xx}-C_{yy}$ corresponds to the Bell variable, constructed from the quark pair spin-correlation matrix elements $C_{xx}$ and $C_{yy}$~\cite{Cheng:2025cuv}.

Belle is an energy-asymmetric $e^+e^-$ collider, so the produced $\Upsilon(2S)$ carries a longitudinal boost along the beam axis in the laboratory frame with velocity $\beta=0.3\sim 0.4$, depending on the beam energies at Belle or Belle II~\cite{Aushev:2010bq,Belle-II:2018jsg}. The transformation of the $\chi_{b2}$ kinematics from the c.m.~frame to the laboratory frame is therefore described by a Lorentz boost  $\Lambda(\beta)$. Under this boost, the helicity state of the $\chi_{b2}$ undergoes a Wigner rotation associated with its little group, corresponding to a rotation of $\theta_\chi$ by an angle $\theta_\beta\in[0,\pi]$, given by~\cite{Yu:2021zmw}
\beq
\cos\theta_\beta=\frac{v_{\chi}+\beta \cos\theta_\chi}{\sqrt{(1+\beta v_{\chi}\cos\theta_\chi)^2-(1-\beta^2)(1-v_\chi^2)}},
\eeq
where $v_\chi$  denotes the speed of $\chi_{b2}$ in the c.m.~frame. Since the radiative transition $\Upsilon(2S)\to\gamma \, \chi_{b2}$ involves only a small recoil momentum, the $\chi_{b2}$ is nonrelativistic in the c.m.~frame and satisfies $v_\chi\ll 1$. 
In this limit, the dependence of the $\theta_\beta$ on the velocity $\beta$ cancels, and the Wigner rotation angle reduces to $\theta_\beta\simeq \theta_\chi$, so the Lorentz boost aligns the momentum of the $\chi_{b2}$ nearly collinearly with that of the parent $\Upsilon(2S)$ in the laboratory frame. 
As a consequence, the differential distribution in Eq.~\eqref{eq:fac} becomes independent of $\theta_\chi$ and $\phi_p$ in the laboratory frame. The Artru-Collins asymmetry can then be extracted experimentally from the average value of $\cos(\phi_1+\phi_2)$,
\begin{widetext}
\begin{align}
A_{12}\equiv 2 \left \langle \cos(\phi_1+\phi_2)\right\rangle
=\frac{1}{2}\frac{\rho_8(m_b)\mathcal{B}\sum_q\mathcal{C}_qH_1^{\sphericalangle, q}(z_1,M_1)H_1^{\sphericalangle,\bar q}(z_2,M_2)}{\rho_8(m_b)\sum_q\mathcal{C}_qD_1^q(z_1,M_1)D_1^{\bar q}(z_2,M_2)+m_b^2\mathcal{C}_gD_1^g(z_1,M_1)D_1^g(z_2,M_2)}.
\label{eq:A12}
\end{align}
\end{widetext}
At LO, the coefficients $\mathcal{B}$ and $\mathcal{C}_{q/g}$ in the laboratory frame are given by
\begin{align}
\mathcal{B}&=\frac{42\sin^2\theta_p}{21\cos^2\theta_p+73},\nn\\
\mathcal{C}_q&=\frac{\alpha_s^2}{480 m_{\chi}^2 \Gamma_{\chi}}(21\cos^2\theta_p+73),\nn\\
\mathcal{C}_g&=\frac{8\alpha_s^2}{15m_{\chi}^4 \Gamma_{\chi}}(\cos^2\theta_p+1),
\end{align}
where $\Gamma_\chi$ denotes the total width of the $\chi_{b2}$. The $\sin\theta_p$ dependence of $\mathcal{B}$ maximizes the asymmetry in the central region, thereby enhancing the observable spin effect. By contrast, in the c.m.~frame the explicit dependence of these coefficients on $\theta_\chi$ and $\phi_p$ causes cancellations over phase space, suppressing the integrated signal. The asymmetric beam energies at Belle remove this angular dependence in the laboratory frame, allowing the asymmetry to survive integration and providing a clear advantage over energy-symmetric colliders. 

Beyond the LO picture described above, higher-order QCD corrections modify the hard coefficients and induce CS and CO mixing, thereby blurring the simple correspondence between the CS $gg$ and CO $q\bar{q}$ channels. However, up to next-to-leading order (NLO) in QCD, because of the vanishing of the amplitude $b\bar{b}(^3S_1^{[8]})\to gg$, the asymmetry still originates predominantly from the light-quark pair fragmentation in the large-thrust two-jet region, where the hard multijet radiation contribution is largely suppressed. By analyzing the Feynman diagrams at the one-loop level, it is straightforward to infer that the $q\bar{q}$ pair produced in the CS $b\bar{b}(^3P_2^{[1]})\to q\bar{q}g$ channel has the same color-octet structure as that produced in the CO channel.   We therefore expect the qualitative sensitivity of $A_{12}$ to the CO  contribution to remain stable. In such a case, the NRQCD factorization scale dependence of the SDCs for the CS $b\bar{b}(^3P_2^{[1]})\to q\bar{q}g$ channel will be canceled by the CO LDME. Furthermore, the impact of NLO QCD corrections is expected to be reduced due to the partial cancellations in the normalized asymmetry, and a dedicated NLO calculation is needed for a precision extraction of $\rho_8(m_b)$.

\vspace{3mm}
\emph{Numerical results and discussion---}
\begin{figure}
    \centering
    \includegraphics[width=\linewidth]{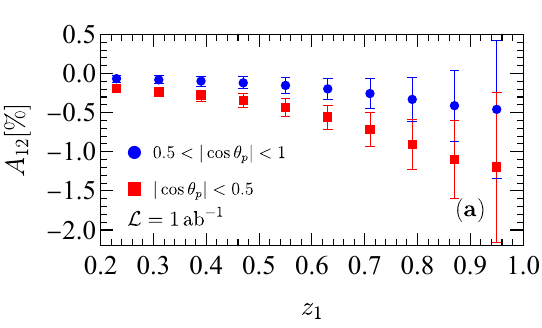}
    \includegraphics[width=\linewidth]{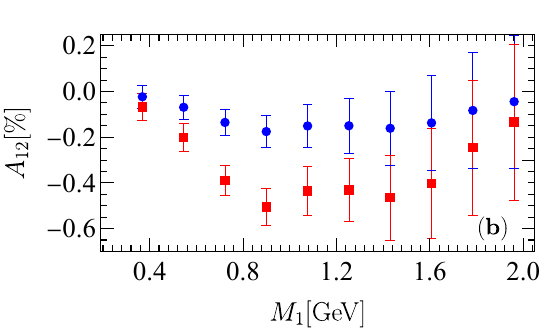}
    \caption{Artru-Collins asymmetry $A_{12}$ in $\chi_{b2}$ decay with the $\chi_{b2}$ produced in the laboratory frame at an integrated luminosity $\mathcal{L}=1~{\rm ab}^{-1}$, shown as a function of $z_1$ (a) and $M_1$ (b) for two bins in $\cos\theta_p$, with all other kinematical variables integrated over.}
    \label{fig:A12}
\end{figure}
With the method developed above, the Artru-Collins asymmetry in $\chi_{b2}$ decay, with the $\chi_{b2}$ produced in the laboratory frame, can be directly estimated. We evaluate this asymmetry using the lattice central value $\rho_8(m_b)=0.044$, defined at the scale $m_b\simeq 4.6~{\rm GeV}$~\cite{Bodwin:2007zf}. The statistical uncertainty of $A_{12}$ in Eq.~\eqref{eq:A12} can be estimated as
\beq
\delta A=\sqrt{\frac{2-A_{12}^2}{N}}\simeq \sqrt{\frac{2}{N}},
\eeq
where $N$ denotes the number of selected events after the kinematic cuts and the approximation holds for $A_{12}\ll 1$.
To assess the statistical sensitivity to $\rho_8(m_b)$, the following projections include statistical uncertainties only, while the impact of the DiFF uncertainties is considered separately.

To quantify their impact on $A_{12}$, we use the JAM global-fit results~\cite{Pitonyak:2023gjx,Cocuzza:2023oam,Cocuzza:2023vqs}. The unpolarized quark DiFFs are typically constrained at the $\lesssim5\%$ level, while the uncertainties of $H_1^{\sphericalangle,q}$ are below approximately $20\%$ over most of the relevant phase space. The dominant uncertainty arises from the unpolarized gluon DiFF $D_1^g$, whose current uncertainty is about $100\%$, mainly due to its indirect determination through QCD evolution~\cite{Cocuzza:2023vqs,Mahaut:2025hie}. Propagating the current DiFF uncertainties to $A_{12}$ yields a relative uncertainty of about $50\%$--$60\%$, whereas the quark DiFFs alone contribute $10\%$--$20\%$, depending on the kinematics. An improved determination of $D_1^g$ is therefore essential for a precision extraction of $\rho_8(m_b)$ and could benefit from dedicated gluon-sensitive dihadron measurements, such as $\chi_{b0}$ decays at Belle II~\cite{He:2026aeh}. With such improvements, the DiFF uncertainties are expected to become subleading relative to the projected statistical uncertainty.

The numerical results for $A_{12}$ are shown in Fig.~\ref{fig:A12} as functions of $z_1$ and $M_1$ for two polar angle bins, $0.5<|\cos\theta_p|<1$ (blue) and $|\cos\theta_p|<0.5$ (red), assuming an integrated luminosity of $\mathcal{L}=1~{\rm ab^{-1}}$, with all other kinematical variables integrated over. 
This binning explicitly demonstrates the enhancement of the asymmetry in the central region. The dependence on $z_1$ and $M_1$ exhibits characteristic features consistent with existing observations of the Artru-Collins asymmetry in $\pi^+\pi^-$-dihadron pair production at Belle~\cite{Belle:2011cur,Courtoy:2012ry,Cocuzza:2023vqs}. 
The typical magnitude of the asymmetry reaches the percent level. Although this is smaller than that measured in the $\pi^+\pi^-$-dihadron pair production at  Belle~\cite{Belle:2011cur} due to dilution from gluon channel, it remains well within the sensitivity of Belle.
For comparison, the CLEO extraction $\rho_8(m_b)=0.160$ from measurements of $\chi_{bJ}(1P)\to D^0X$~\cite{CLEO:2008bsq} enhances the relative CO
$q\bar q$ contribution in Eq.~\eqref{eq:A12},  increasing the magnitude of $A_{12}$ relative to the lattice benchmark while leaving the
shapes of its $z_1$ and $M_1$ dependences essentially unchanged. In addition, the $z_1$ and $M_1$ dependences are not expected to exhibit the sharp endpoint
enhancement discussed in Ref.~\cite{Bodwin:2007zf} since the relevant partonic momentum fraction is integrated over in forming
$A_{12}$. Any residual endpoint effect is part of the missing NLO corrections.

\begin{figure}
    \centering
    \includegraphics[width=\linewidth]{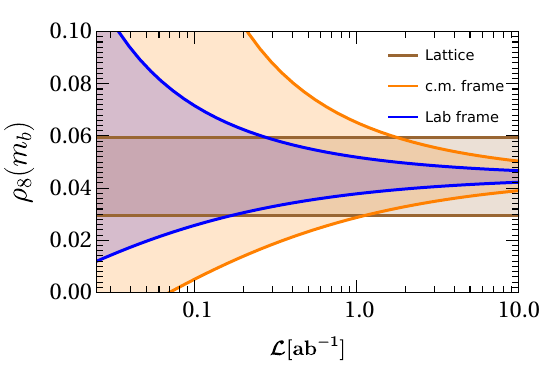}
    \caption{Projected statistical sensitivity to $\rho_8(m_b)$ from the Artru-Collins asymmetry in $\chi_{b2}$ decay at Belle II as a function of the integrated luminosity, for $\chi_{b2}$ produced in the laboratory frame (blue band) and in the c.m.~frame (orange band).}
    \label{fig:rho8}
\end{figure}

To quantify the sensitivity to $\rho_8(m_b)$, we perform a binned analysis in $\theta_p$, exploiting the enhancement of the asymmetry in the central angular region.
Owing to the limited event statistics, the remaining kinematic variables are integrated over, with $z_i\in[0.19,0.99]$ and $M_i\in[0.28,2.05]~{\rm GeV}$~\cite{Cocuzza:2023vqs}.
The projected constraints are obtained from a $\chi^2$ analysis.  
The resulting 68\% confidence level sensitivity to $\rho_8(m_b)$ in the laboratory frame as a function of integrated luminosity is shown by the blue band in Fig.~\ref{fig:rho8}, while the corresponding sensitivity in the c.m.~frame is indicated by the orange band. 
In the c.m.~frame, applying the same binning strategy as in the laboratory frame, namely retaining only $\theta_p$ and integrating over $\phi_p$ and $\theta_\chi$, causes the asymmetry to vanish due to complete phase space cancellations. We therefore bin instead in the production angle $\theta_\chi$ for comparison. However, residual cancellations persist after integrating over $\theta_p$ and $\phi_p$, substantially suppressing the asymmetry and yielding significantly poorer sensitivity than in the laboratory frame.  
For reference, the lattice result, $\rho_8(m_b)=0.044\pm 0.015$~\cite{Bodwin:2007zf}, is shown as the brown band, while the CLEO extraction, $\rho_8(m_b)=0.16^{+0.071}_{-0.047}$~\cite{CLEO:2008bsq}, lies far above the lattice value and is therefore not displayed in the same figure. 
The projected precision in the laboratory frame surpasses the current lattice uncertainty with an integrated luminosity of $\mathcal{O}(0.1)~{\rm ab}^{-1}$ and could reach the few-percent level at $\mathcal{L}\sim 10~{\rm ab}^{-1}$. This opens an opportunity to resolve the long-standing discrepancy in bottomonium between lattice  NRQCD calculations and phenomenological extractions of $\rho_8(m_b)$. 
A similar analysis can also be performed for the radiative transitions $\Upsilon(3S)\to\gamma \chi_{b2}(2P)$, which provides a nontrivial cross-check in view of the expected weak dependence of $\rho_8(m_b)$ on the radial quantum number~\cite{Brambilla:2001xy} and will be investigated in future work.

\vspace{3mm}
\emph{Conclusion---}
In this Letter, we have proposed a novel and clean observable to directly probe the color-octet mechanism in NRQCD through the Artru-Collins asymmetry in hadronic decays of the $P$-wave bottomonium state $\chi_{b2}$. From NRQCD power counting, $P$-wave quarkonia provide an ideal laboratory for this study since the color-singlet  and  color-octet contributions enter the decay rates at the same order, whereas, at LO, the spin-dependent asymmetry arises exclusively from the color-octet channel.  This distinctive feature makes the asymmetry a direct and unambiguous signal of color-octet dynamics.  A central result of our analysis is that the asymmetric beam energies at Belle preserve nontrivial transverse spin azimuthal correlations between the quark and antiquark in $\chi_{b2}$ decay in the laboratory frame, thereby avoiding the strong phase space cancellations that suppress the
effect in the center-of-mass frame. By employing DiFFs from global fits, we have demonstrated that this observable enables a precise determination of the ratio of NRQCD long-distance matrix elements, $\rho_8(m_b)$. The projected statistical sensitivity surpasses current lattice QCD uncertainty with an integrated luminosity of $\mathcal{O}(0.1)~{\rm ab}^{-1}$ and could reach the few-percent level at Belle II, thereby providing an opportunity to resolve this long-standing discrepancy.
Our results establish spin correlation in $P$-wave quarkonium decays as a powerful and complementary framework for tests of NRQCD.

\vspace{3mm}
\emph{Acknowledgments---}
We thank Y. Jia, B. A. Kniehl and C. P. Shen for helpful comments on the manuscript. This work is partly supported by the National Natural Science Foundation of China under Grants No.~12547174, No.~12422506, and No.~12342502; CAS under Grant No.~E429A6M1; the China Postdoctoral Science Foundation under Grant No.~2026M793746; and the Fundamental Research Funds for the Central Universities under Grant No. buctrc202432. The authors gratefully acknowledge the valuable discussions and insights provided by the members of the Collaboration on Precision Tests and New Physics (CPTNP).

\bibliographystyle{apsrev}
\bibliography{reference}

\end{document}